\documentstyle[epsfig]{aipproc}

\begin{document}

\title{A brief introduction to Luttinger liquids
}
\author{Johannes Voit}
\address{Theoretische Physik 1, Universit\"{a}t Bayreuth, 
D-95440 Bayreuth (Germany) 
}
\maketitle
\begin{abstract}
I give a brief introduction to Luttinger liquids. Luttinger liquids are
paramagnetic one-dimensional metals without Landau quasi-particle excitations.
The elementary excitations are collective charge and spin modes, leading
to charge-spin separation. Correlation functions exhibit power-law behavior.
All physical properties can be calculated, e.g. by bosonization, and depend
on three parameters only: the renormalized coupling constant $K_{\rho}$, and
the charge and spin velocities. I also discuss the stability of Luttinger
liquids with respect to temperature, interchain coupling, lattice effects
and phonons, and list important open problems.
\end{abstract}


\section*{What is a Luttinger liquid anyway?}
\label{moti}
Ordinary, three-dimensional metals are described by Fermi liquid theory.
Fermi liquid theory is about the importance of electron-electron interactions
in metals. It states that there is a 1:1-correspondence between the 
low-energy excitations of a free Fermi gas, and those of an interacting 
electron liquid which are termed ``quasi-particles'' \cite{fermi}. Roughly
speaking, the combination of the Pauli principle with low excitation
energy (e.g. $T \ll E_F$) and the large phase space available in 3D, produces
a \em very dilute \rm
gas of excitations where interactions are sufficiently harmless so
as to preserve the correspondence to the free-electron excitations.
Three key elements are: (i) The elementary excitations of the Fermi liquid
are quasi-particles. They lead to a pole structure (with residue $Z$ --
the overlap of a Fermi surface electron with free electrons) in
the electronic Green's function which can be -- and has been -- observed
by photoemission spectroscopy \cite{claess}. (ii) Transport is described
by the Boltzmann equation which, in favorable cases, can be quantitatively
linked to the photoemission response \cite{claess}. 
(iii) The low-energy physics is 
parameterized by a set of Landau parameters $F_{s,a}^{\ell}$
which contain the residual
interaction effects in the angular momentum charge and spin channels. \em
The correlations in the electron system are weak, although the interactions
may be very strong. \rm 

Fermi liquid theory breaks down for one-dimensional (1D) metals.
Technically, this happens because some vertices Fermi liquid theory assumes
finite (those involving a $2k_F$ momentum transfer) actually diverge because
of the Peierls effect. An equivalent intuitive argument is that in 1D,
perturbation
theory never can work even for arbitrarily small but finite interactions:
when degenerate perturbation theory is applied to the coupling of the 
all-important electron states at the Fermi \em points \rm $\pm k_F$, it will
split them and therefore remove the entire Fermi surface! A free-electron-like
metal will therefore not be stable in 1D. The underlying physical picture
is that the coupling of quasi-particles to collective excitations is small
in 3D but large in 1D, no matter how small the interaction. \em Correlations
are strong even for weak interactions! \rm

1D metals are described as Luttinger liquids \cite{Haldane,myrev}. 
\em A Luttinger liquid is a paramagnetic one-dimensional metal without
Landau quasi-particle excitations. \rm ``Paramagnetic'' and ``metal'' require
that the spin and charge excitations are gapless, more precisely with
dispersions
$\omega_{\nu} \approx v_{\nu} |q|$ ($\nu=\rho,\sigma$ for charge and spin). 
Only when this requirement is fulfilled,
a Luttinger liquid can form. 
The charge and spin modes (holons and spinons)
possess different excitation energies $v_{\rho}
\neq v_{\sigma}$ and are bosons. 
This leads to the separation of charge and spin of an
electron (or hole) added to the Fermi sea, in space-time, or $q-\omega$-space.
Charge-spin separation prohibits quasi-particles: The 
pole structure of the Green's function is changed
to branch cuts, and therefore the quasi-particle residue $Z$ is zero. 
Charge-spin separation in space-time can be nicely observed in computer
simulations \cite{jagla}.

The bosonic nature of charge and spin excitations, 
together with the reduced dimensionality leads to a peculiar kind of 
short-range order at $T=0$. The system is at a (quantum) critical point,
with power-law correlations,
and the scaling relations between the exponents of its correlation functions
are parameterized by renormalized coupling constants $K_{\nu}$. 
The individual exponents are non-universal, i.e.  depend
on the interactions. For
Luttinger liquids, $K_{\nu}$ is the equivalent of the Landau parameters.
As an example, the momentum distribution function 
$n(k) \sim (k_F-k)^{\alpha}$ for $k \approx k_F$ with $\alpha = (K_{\rho}
+ K_{\rho}^{-1}-2)/4$. This directly illustrates the absence of 
quasi-particles: In a Fermi liquid, $n(k)$ has a jump at $k_F$ with
amplitude $Z$.

\section*{Bosonization, or how to solve the 1D many- \\
body problem by harmonic 
oscillators}
\label{lemodel}
\begin{figure}[b!] 
\centerline{\epsfig{file=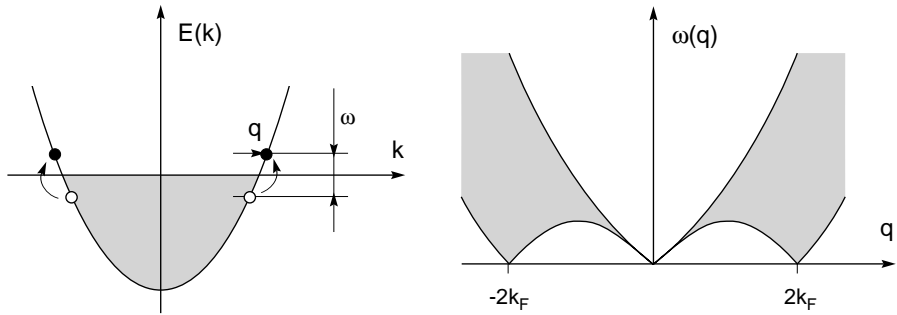,
height=5.0in,
angle=90}}
\vspace{10pt}
\caption{Particle-hole excitations in 1D (left). The spectrum of allowed 
excitation has no low-energy states with $0 \leq |q| \leq 2k_F$.}
\label{fig1}
\end{figure}

The appearance of charge and spin modes as stable, elementary excitations 
in 1D fermion systems can
be rationalized from the spectrum of allowed particle-hole excitations. In
1D,  low-energy particle-hole pairs 
with momenta between $0$ and $2k_F$ are not
allowed, and for $q \rightarrow 0$, the range of allowed 
excitations shrinks to a
one-parameter spectrum $\omega_{\nu} \approx v_{\nu} |q|$, 
indicating stable particles (cf. Fig.\ \ref{fig1}).
True bosons are then obtained as linear combinations of these
particle-hole excitations with a definite momentum $q$. Most importantly,
we now can rewrite any interacting fermion Hamiltonian, 
provided its charge and spin excitations are gapless, as a harmonic 
oscillator and 
find an operator identity allowing to express fermion operators as functions
of these bosons. This is the complete bosonization program.

For free fermions, the Hamiltonian describing the
\em excitations \rm out of the ground state (the Fermi sea), a
can be expressed as a bilinear
in the bosons, 
\begin{equation}
H = \sum_{\nu = \rho, \sigma} \sum_q v_{\nu} |q| \left( b_{\nu,q}^{\dag}
b_{\nu,q} + 1/2 \right) \;,
\end{equation}
with $v_{\nu} = v_F$, the Fermi velocity.
Both the spectrum and the multiplicities of the 
states, i.e. the Hilbert space, of the fermion and boson forms are identical
\cite{Haldane}. 

What happens in the presence of interactions? One possibility is that the
interactions open a gap in the spin and/or charge excitation spectrum. The
system then no longer is paramagnetic and/or metallic.
With a charge gap, we have a 1D Mott 
insulator, with a spin gap a conducting system with strong charge density
wave or superconducting correlations, and gaps in both channels imply a
band insulator. Luttinger liquid theory cannot be applied anymore. 
In the other case, charge and spin excitations remain gapless: a
Luttinger liquid is formed. Then, electron-electrons interactions will
make $v_{\sigma} \neq v_{\rho} \neq v_F$, leading to charge-spin separation. 
Interactions will also renormalize
the electronic compressibility and magnetic susceptibility, and the charge
and spin stiffnesses, and by comparing the velocities measuring this
renormalization to $v_{\nu}$, the correlation exponents $K_{\nu}$ can be
defined. The $K_{\nu}$ therefore only depend on the low-energy properties
of the Hamiltonian. Two parameters per degree of freedom, $K_{\nu}$ and
$v_{\nu}$, completely describe the physics of a Luttinger liquid.

From model studies, e.g. on the 1D Hubbard model \cite{hjs}
and related models \cite{myrev}, the following picture emerges:
(i) $K_{\nu}=1$ describes free electrons, and 
$K_{\sigma}=1$ is required by spin-rotation invariance. (ii) 
$K_{\rho} > 1$ for effectively attractive interactions, and $K_{\rho}<1$
for repulsive interactions. (iii) For the 1D Hubbard model, 
$K_{\rho}$ decreases from $1$ to $1/2$ as the electron
repulsion $U$ varies between $0 \dots \infty$. (iv) $K_{\rho}<1$ decreases
with increasing interaction range. For any finite range, there is a 
characteristic minimal $K_{\rho}$, which approaches zero, as the interaction
range extends to infinity. (v) $v_{\sigma} \leq v_F$ for repulsive 
interactions. $v_{\sigma}$ measures the magnetic exchange $J$. (vi)
$v_{\rho} > v_F$ for repulsive interactions, and the more so the longer
the interaction range. In the limit of unscreened Coulomb interaction,
$v_{\rho} \rightarrow \infty$, and the charge fluctuations then become the
1D plasmons \cite{wicrys} with $\omega_{\rho}(q) \propto |q| {\ln|q|}$. 
(vii) Electron-phonon interaction decreases the $v_{\nu}$, and most often
also $K_{\rho}$. Interaction with high-frequency dispersionless molecular
vibrations can enhance $K_{\rho}$ and lead to superconductivity \cite{phon}.

To complete our bosonization program, a local fermion operator must be 
expressed in terms of bosons. Exact operator identities are available
for the Luttinger model \cite{Haldane,myrev} which can be summarized 
schematically as
\begin{equation}
\Psi_s (x) \sim \exp \left\{ i \sum_{\nu} \sum_p e^{ipx} ( \ldots )
\left( b_{\nu,p} + b_{\nu,-p}^{\dag} \right) \right\}\;.
\end{equation}
This fermion-boson transformation turns bosonization into a useful device:
all correlation functions can  be calculated as simple harmonic oscillator
averages, repeatedly using the two important identities $e^A e^B = 
e^{A+B} e^{[A,B]/2}$ for $[A,B]$ a complex number,
and $\langle e^A \rangle = \exp(\langle A^2 \rangle / 2)$ valid for 
harmonic oscillator expectation values. 
As a consequence, \em Luttinger liquid predictions for all physical 
properties can be produced. \rm Examples are given in the next section.
[The behavior of the momentum distribution
function $n(k)$ discussed above, has been obtained from the single-particle 
Green's function $\langle T \Psi(xt) \Psi^{\dag}(00) \rangle$ in precisely
this way].

Bosonization is an easy and transparent way to calculate the properties
of Luttinger liquids. However, it is not the only method. More general,
and more powerful is the direct application of conformal field theory
to a microscopic model of interacting fermions. For Luttinger liquids,
both methods become identical, and one might view 
bosonization as solid state physicist's way
of doing conformal field theory. Also Green's functions methods have been
used successfully.

\section*{Predictions for experiments}
\label{appl}
This section summarizes some important properties of
Luttinger liquids in the form of experimental predictions. The underlying
theoretical correlation functions can be found elsewhere \cite{myrev}. We
discuss a single-band Luttinger liquid. For multiband systems, such as the
metallic carbon nanotubes, the exponents differ from those give here, but
can be calculated in the same way \cite{tubeexp}.

The thermodynamics is not qualitatively different from a Fermi liquid,
with a linear-in-$T$ specific heat (expected both for 1D fermions and 
bosons!), and $T$-independent Pauli susceptibility and electronic 
compressibility
\begin{equation}
C(T) = \frac{1}{2}\left( \frac{v_F}{v_{\rho}} + \frac{v_F}{v_{\sigma}}
\right) \gamma_0 T \;, \; \; 
\chi = \frac{2 K_{\sigma}}{\pi v_{\sigma}} \;,\; \;
\kappa = \frac{2 K_{\rho}}{\pi v_{\rho}} \;.
\end{equation}
More interesting are the charge and spin correlations at 
wavenumber multiples of $k_F$ which display the $K_{\rho}$-dependent 
power laws discussed above. In the electronic structure factor $S(k)$ and 
NMR spin-lattice relaxation rate $T_1^{-1}$, they translate into
\begin{equation}
\label{dens}
S(k) \sim |k-2k_F|^{K_{\rho}} + |k-4k_F|^{4 K_{\rho}-1} \;, \;\;
T_1^{-1} \sim T + T^{K_{\rho}} \;.
\end{equation}
The structure factor can be interpreted as showing fluctuations both of
Peierls-type ($2k_F$) and of Wigner-crystal-type ($4k_F$) charge density
waves, and the two terms in $T_1^{-1}$ come from the $q \approx 0$ and
$2k_F$ spin fluctuations. Evidence for such behavior has been found, e.g.
in TTF-TCNQ  \cite{pouget} for $S(k)$, and $(TMTSF)_2 ClO_4$ \cite{creuzet} 
for $T_1^{-1}$.
Transport properties depend on the scattering mechanisms assumed. If we
consider electron-electron scattering in a band with filling factor $1/n$,
we obtain from the current-current correlations \cite{giam}
\begin{equation}
\label{conduc}
\rho(T) \sim T^{n^2 K_{\rho}-3} \;, \;\; \sigma(\omega) 
\sim \omega^{n^2 K_{\rho} -5 } \;.
\end{equation}
The second law has apparently been observed in salts based on TMTSF
\cite{schwartz}. These predictions ideally give information on the 
power-law behavior of correlations, and on the underlying value of $K_{\rho}$,
which, of course, must be the same for different experiments in any specific
material.

\begin{figure}[b!] 
\centerline{\epsfig{file=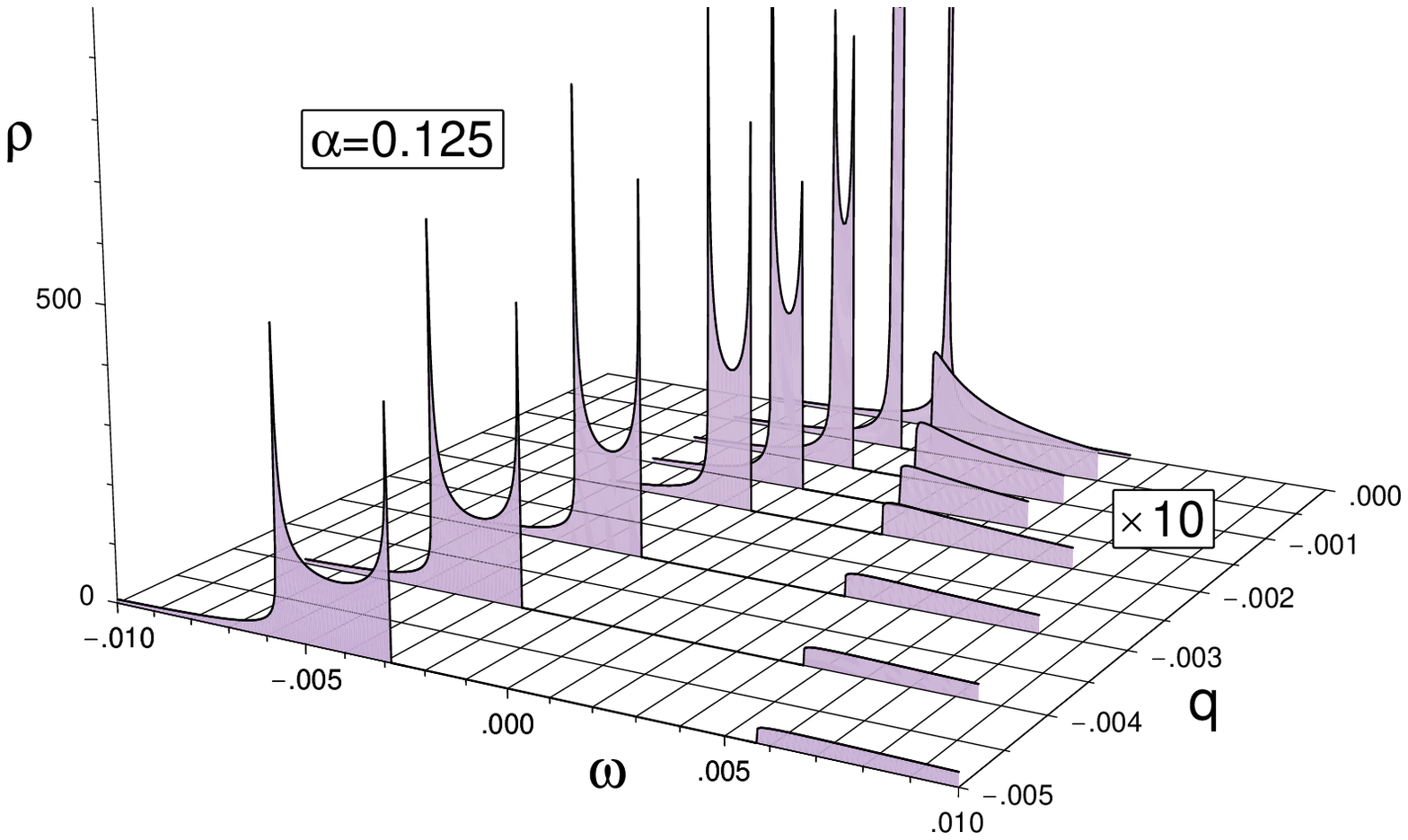,height=1.9in} \hspace{-24pt}
\epsfig{file=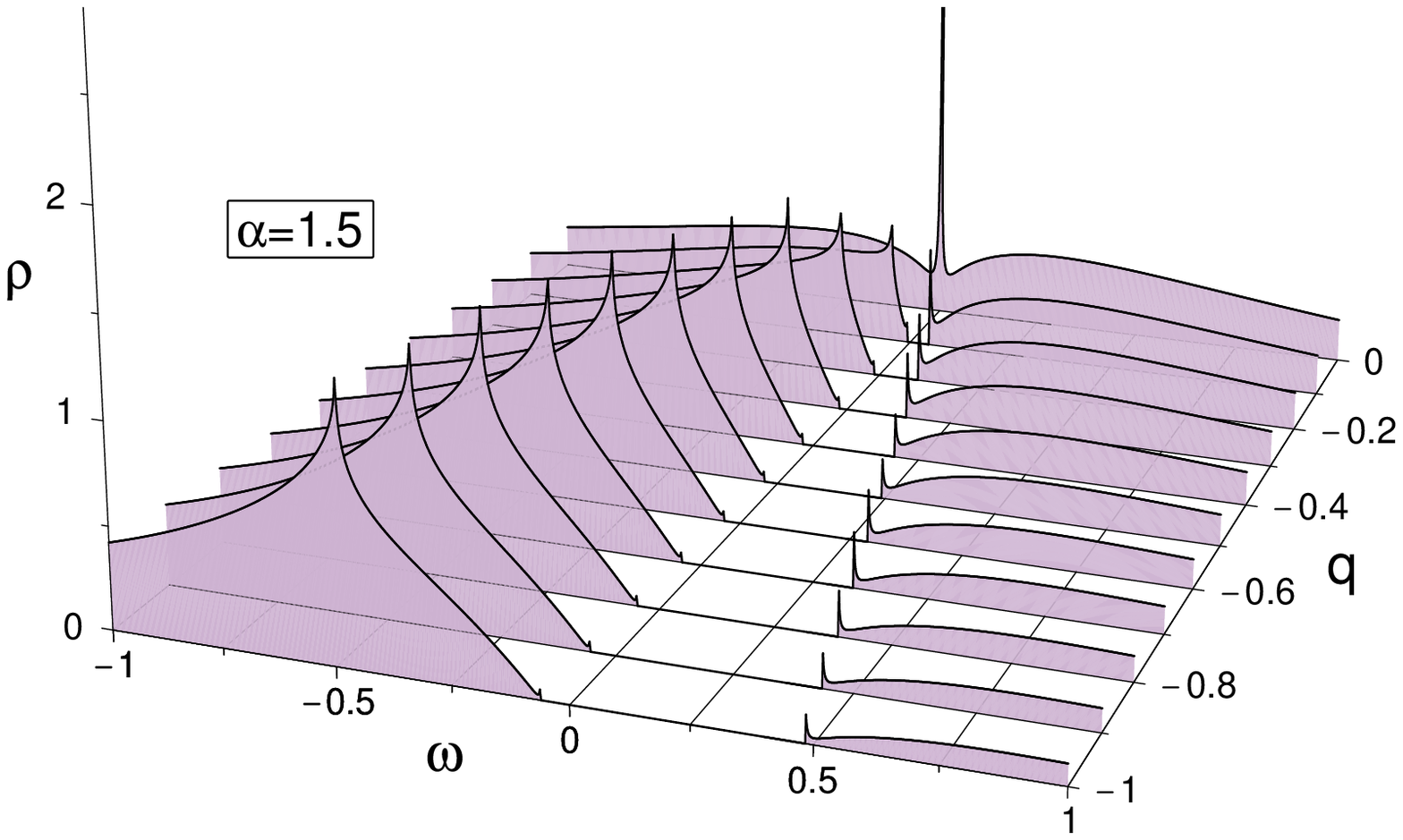,height=1.9in}}
\vspace{10pt}
\caption{Spectral functions of a Luttinger liquid. The three signals represent
the holon, the spinon, and the shadow bands (left to right). 
Left panel: weak/short-range 
interactions, $\alpha=1/8$ ($K_{\rho}=1/2$). Right panel: strong/long-range 
interactions, $\alpha=1.5$ ($K_{\rho}=1/8$). }
\label{fig2}
\end{figure}

In order to see charge-spin separation, one must perform $q$- and 
$\omega$-resolved spectroscopy (or time-of-flight measurements). 
Photoemission spectroscopy is the first choice because it directly probes
single-particle excitations \cite{mgjv}. 
With some approximations, it measures the
imaginary part of the electronic Green's function, and Luttinger liquid
theory predicts, cf. Fig.\ \ref{fig2} \cite{myspec,jphys}
\begin{equation}
\rho(q,\omega) = \frac{-1}{\pi} {\rm Im} G(q+k_F, \omega+E_F)
\sim (\omega - v_{\sigma}q)^{\alpha - 1/2} |\omega- v_{\rho}q|^{(\alpha-1)/2}
(\omega + v_{\rho} q)^{\alpha/2} \; .
\end{equation}
One finds \em two dispersing singularities \rm (with interaction dependent
exponents; for $\alpha$, cf. above) which demonstrates
that the electron ejected from the material is composed out of two more
elementary excitations. 
By $q$-integration, one can obtain the density of
states, $N(\omega) \sim |\omega|^{\alpha}$, and by $\omega<0$-integration
$n(k)$.  A practical comment: the easy part is the 
calculation of the Green's function in bosonization. The difficult part
is the Fourier transformation if, e.g., the result must satisfy sum rules.
Charge-spin separation is also visible, though with different exponents,
in dynamical density and spin correlation at $q \approx 2k_F$ \cite{jphys}.
They can
be measured, in principle, by EELS, inelastic neutron scattering, or
Raman scattering. 

\section*{Stability of Luttinger liquids}
Luttinger liquid theory crucially relies on one-dimensionality. Moreover,
most of our discussion was for $T=0$, and ignored phonons, lattice effects,
impurities, etc. Are these factors detrimental to Luttinger liquids? In many
cases, the answer will depend on the scales one considers.

Finite temperature is not a problem, and the correlation functions discussed
above can be calculated for $T>0$. Quite generally, however,
divergences will be
cut off by $T$ whenever $T > \omega, v_{\nu}|q|, \ldots$. Also charge-spin
separation will be masked in the spectral function when $(v_{\rho}-
v_{\sigma}) q < T$ \cite{nakam}. 

Interchain tunneling will introduce 3D effects. 
Depending on the on-chain interactions, it will either produce a crossover
to a Fermi liquid (weak interactions), or to a long-range ordered 3D
insulating or superconducting phase (strong interactions) \cite{ichain}.
In any event, \em a Luttinger liquid is unstable towards 3D coupling at
low enough temperature (scales). \rm However, on high enough scales, it
will be unaffected by 3D coupling, and coming from 
there, one will encounter a 
crossover temperature below which 3D correlations will build up, and 1D
physics will be strongly modified. At still lower temperature, a
phase transition may take place into a long-range ordered 3D state. When
going to a Fermi liquid, the crossover is gradual, and Luttinger-like
spectral functions can be observed somewhat off the Fermi energy
\cite{medenthes}.

Other sources of concern are phonons and lattice effects. Various studies
of phonons coupled to Luttinger liquids have shown that depending on 
details of the electron-electron and electron-phonon interactions, a 
Luttinger liquid may remain stable, though renormalized, when phonons are
added \cite{phon}. Alternatively, the electron-phonon interaction could
lead to the opening of a spin gap, 
and thereby destabilize the Luttinger 
liquid. This situation is described by a different model due to Luther
and Emery, but the correlation functions continue to carry certain 
remnants of Luttinger physics, like non-universal power laws stemming from
the gapless charges (the system remains conducting), and charge-spin
separation \cite{zittartz}. 

When the crystal lattice is important (commensurate band filling), the
system may become insulating. For a 1D band insulator, Luttinger liquid 
physics is expected to be lost completely, although not much is known 
firmly \cite{vescoli}. More interesting is the case of a Mott insulator,
brought about by electronic correlations. However even here, charge-spin
separation still is seen, e.g. in photoemission both in theory 
\cite{zittartz} and in experiments on SrCuO$_2$ \cite{kim}. Moreover,
far above the (charge or spin) gaps, they should no longer influence 
the physics, and genuine Luttinger liquid behavior is expected there.

\section*{How to apply Luttinger liquid theory?}
We discuss the example of the organic conductor TTF-TCNQ, starting from
a recent photoemission study \cite{zwick}. Photoemission shows a valence
band signal whose dispersion is in qualitative agreement with a simple
H\"{u}ckel band structure, along the 1D chains, and no dispersion 
perpendicular. Two discrepancies between the data and a quasi-particle 
picture can be resolved in a Luttinger liquid picture: (i) the experimental
dispersions are bigger than those expected from the density of states at
the Fermi level within the H\"{u}ckel band structure. 
(ii) The lineshapes are anomalous
in that the signal on the TTF band has a tail reaching up to $E_F$ at all
$k$, while it has a low-energy shoulder with little dispersion on TCNQ. 
Both findings are consistent with Luttinger liquid spectral functions,
with $K_{\rho} \ll 1/2$ on TTF, and with $1/2<K_{\rho}<1$ on TCNQ. Also
recall that Luttinger liquids show more dispersion than Fermi liquids
because of the upward renormalization of $v_{\rho}$ by interactions.

Is this assignment consistent with other information? It is consistent for
the TTF band. In fact, the magnetic susceptibility is rather independent
of temperature \cite{taka}, and diffuse X-ray scattering observes strong
$4k_F$ density fluctuations at high temperatures \cite{pouget}. It is not
consistent, however, for the TCNQ band where the susceptibility is strongly
$T$-dependent, with an activated shape. This can be taken as an indication
of a spin gap, and suggests that the Luther-Emery model might be a better
choice. The observation of $2k_F$ density fluctuations on TCNQ is consistent
with both assignments, and the spectral function of the Luther-Emery model
can at least qualitatively describe the data.

Evidence for or against such hypotheses must come from further experiments.
Optics shows a far-IR pseudogap \cite{ttfopt}. 
However, the consistency of the
mid-IR conductivity with (\ref{conduc}) should be checked. Also notice
that \cite{cooper} $\rho(T) \sim T$. One might look into the 
temperature dependence of the spin conductivity in view of theories
discussing the manifestation of charge-spin separation in transport 
properties \cite{csstr}. NMR could look for the $T^{K_{\rho}}$-term of
Eq.\ (\ref{dens}), and Raman scattering could show if the values of 
$v_{\nu}$ measured through two-particle excitations superpose to the
dispersions of the photoemission peaks. If successful, the Luttinger liquid
theory will provide a consistent phenomenology for the low-energy properties
of this material, and will have predictive power for future experiments.

\section*{Aspects of mesoscopic systems}
Due to the small sample size, boundary conditions become of importance,
and may dominate the physics. As an example, for a quantum wire, the 
conductance is given by $G_n = 2 n K_{\rho} e^2/h$ where $n$ is the 
number of conducting channels \cite{kafi}. 
When the wire is coupled to Fermi liquid
leads, however, the interaction renormalization is absent \cite{msss}, 
and $G_n = 2 n e^2 / h$ -- a boundary effect! 

The influence of isolated impurities on transport, or tunneling through
quantum point contacts, is an important problem \cite{kafi,bce}. At higher
temperatures (voltages), there will be corrections to the (differential)
conductance $\delta G \sim T^{K_{\rho}-1}$, resp. $\delta(dI/dV) \sim 
V^{K_{\rho}-1}$. 
With repulsive 
interactions and at low energy scales, 
an impurity will cut the quantum wire into two segments
with only a weak link between them. In this case, the conductance, resp.
differential conductance, vary as $G(T) \sim T^{K_{\rho}^{-1}-1}$, 
resp. $dI/dV \sim V^{K_{\rho}^{-1}-1}$. The physical origin of this
effect is the establishment of a strong Friedel oscillation around the
impurity which will increasingly backscatter the electrons at lower energy
scales. 

An impurity can therefore be assimilated with open boundary conditions. 
This identifies the exponents just described as members of a larger
class of \em boundary critical exponents. \rm Quite generally, 
1D interacting fermions with open boundaries and gapless excitations form a 
\em bounded Luttinger liquid \rm state, rather similar to ordinary 
Luttinger liquids but with a different set of exponents and
scaling relations \cite{bce}. 
The $K_{\nu}$ are properties of the Hamiltonian, and therefore 
independent of boundary conditions. The
correlation functions, and their exponents, however depend on 
boundary conditions.  

A particularly nice
experiment demonstrating this relation,
has been performed on carbon nanotubes \cite{bockrath}.
With different preparations, it is possible to tunnel electrons from 
electrodes either into the end of nanotubes, or into their bulk. In the
first case, conductance and differential conductance measure 
the power-laws just described for tunneling through a weak link, while
for tunneling into the bulk, they measure the bulk density of states, 
described in the context of photoemission. The exponents differ slightly
from those given here because of the peculiar band structure of the tubes
and because the electrons tunnel from a Luttinger liquid into a normal metal
\cite{tubeexp}. 
The remarkable result of this work is that the various experiments can
be described in terms of a \em single coupling constant \rm $K_{\rho}
\sim 0.28$.

\section*{Open questions}
The preceding discussion may suggest that one-dimensional fermions are 
completely understood, at least theoretically. However, many important 
questions remain open, both in theory and experiment. I now list a few 
of them. 

One important problem relates to scales. While common folklore states that
Luttinger liquids form on energy scales between the electronic bandwidth
or the typical interaction energy, whichever is smaller, on the high-energy
side, and the 3D crossover temperature on the low-energy side, it is not
known with certainty if all predicted properties can indeed be observed in
that range. Can both power laws and charge-spin separation be observed over
the entire range? Some studies seem to suggest that, in the 1D Hubbard model,
the Green's function power laws may be restricted to smaller scales
\cite{hats}. Are these ranges the same for all correlations, or do they depend
on the specific function considered? Do they depend on the specific 
Hamiltonian considered, e.g. on the interaction strength and range, and how?

Concerning mesoscopic systems, only Luttinger liquids with open boundaries 
are thoroughly characterized. It is conceivable that other boundary  
conditions (Fermi liquid leads, boundary fields or spins, superconductors)
lead to new sets of critical exponents. 

What is the spectral weight associated with Luttinger liquid physics in any
given microscopic model, or in any given experimental system? Can one 
measure, in analogy to the quasi-particle residue $Z$ in Fermi liquids,
the weight of the coherent spin and charge modes, with respect to the
incoherent contributions to the Green's function, or to any other correlation
function? How sure can we be that this weight is sufficiently high, so that
experiments (e.g. photoemission) actually see these excitations, and not just
incoherent contributions or bare high-energy excitations? Is the high-energy
physics, far from the Fermi surface, necessarily non-universal and strongly
material- (model-) dependent, as is often claimed?

In the same way, the interpretation of some experiments, e.g. photoemission,
rests crucially on the appropriateness of simple H\"{u}ckel-type 
bandstructures. However, the materials investigated to date, are very
complex, and there is no guarantee that these methods are appropriate.
There are two ways out. (i) More sophisticated band structure methods 
become more performing as the computer power increases, and should attack
the complex materials of interest here \cite{bc}. (ii) One might also look
at novel structures where extremely simple 1D materials can be produced.
One example for this direction are gold wires deposited on a vicinal 
Si(111)5x1 surface, where photoemission may have detected evidence for
charge-spin separation and Luttinger liquid behavior \cite{baer}. (iii)
In mesoscopic wires, both on semiconductor and tube base, we would love
to have spectroscopic experiments made feasible which probe the dynamics
of the elementary excitations beyond transport. As a first step,
the study of ``noise'' 
might provide interesting insights \cite{ar}.

\section*{Acknowledgements}
I am a Heisenberg fellow of Deutsche Forschungsgemeinschaft, 
and received additional support from DFG under SFB 279-B4 and SPP 1073.
Many of my contributions to this field are the fruit of collaborations with
Diego Kienle, Marco Grioni, Anna Painelli,
and Yupeng Wang. Much of my understanding 
of these matters is due to Heinz-J\"{u}rgen Schulz.

\end{document}